\documentclass[
  journal=jacsat,        
  manuscript=communication,    
  layout=traditional,    
  articletitle=true,     
  etalmode=none,    
  maxauthors=999,         
]{achemso}

\makeatletter
\renewcommand*\acs@maxauthors{999}
\makeatother

\SectionNumbersOn
\usepackage[version=4]{mhchem} 
\usepackage{gensymb}
\usepackage{graphicx}
\usepackage{xcolor}
\usepackage[normalem]{ulem}

\newcommand{\POp}{PO$^{+}$}
\newcommand{\mz}{\textit{m}/\textit{z}}

\author{Sanjana Maheshwari}
\altaffiliation{These authors contributed equally.}
\affiliation[UMD]
{Department of Chemistry and Biochemistry, University of Maryland, College Park, MD USA}

\author{Darya Kisuryna}
\altaffiliation{These authors contributed equally.}
\affiliation[UMD IPST]
{Institute for Physical Science and Technology, University of Maryland, College Park, MD USA}

\author{Leah G. Dodson}
\email{ldodson@umd.edu}
\affiliation[UMD]
{Department of Chemistry and Biochemistry, University of Maryland, College Park, MD USA}

\title[Propylene oxide cation]
    {Charge Transfer from Ammonia Neutralizes Propylene Oxide Cations: Implications for the Astrochemistry of Chiral Molecules}
\abbreviations{GDIT,\POp{},\mz{}}
\keywords{propylene oxide,charge-transfer reaction,ion-trap kinetics,astrochemistry,chirality}

\begin{document}

\begin{tocentry}

    \includegraphics[width=\linewidth]{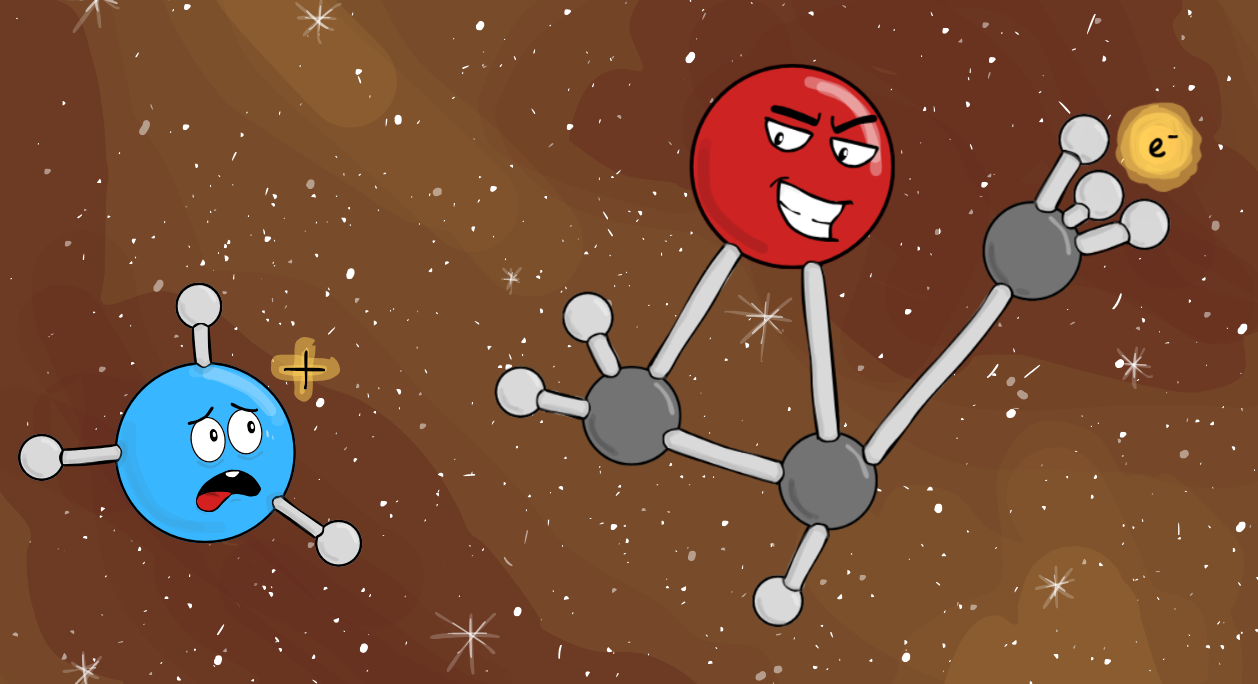}

\end{tocentry}

\begin{abstract}
  To date only one chiral species, propylene oxide, has been observed in the interstellar medium but little is known about the chemistry that leads to a detectable abundance of this molecule in Sagittarius B2. We used a glow-discharge ion source and a room-temperature ion trap to study the neutralization reactions necessary to convert propylene oxide cation (\POp{})---the assumed precursor for propylene oxide in space---into the observed astrochemical. We found that the charge-transfer reaction between \POp{} and ammonia (\ce{NH3}) proceeds with a pressure-independent rate coefficient of $(1.39\pm0.03)\times10^{-12}$ cm$^{3}$ s$^{-1}$ to neutralize \POp{} and form the radical cation \ce{NH3+}. Although this measured rate coefficient is much slower than that predicted by capture theories, the high abundance of \ce{NH3} in Sagittarius B2 motivates the inclusion of this reaction in astrochemical models. We hypothesize that the low ionization energies of many chiral molecules important to origin-of-life theories means these species may exist as cations in the interstellar medium.    
\end{abstract}

\section{Introduction}
Many molecules linked to the origin of life, such as amino acids and sugars, are chiral---making the search for molecules with chiral centers a priority for many astronomers and planetary scientists. While organic precursors to these chiral molecules have been detected in the interstellar medium, molecular clouds, and meteorite samples,\cite{kvenvolden_evidence_1970, rodriguez-almeida_first_2021, morris_cyanoacetylene_1976, lacy1991discovery, weinberger1976circumstellar, brown1975discovery} continued efforts to detect molecules with chiral centers in astrophysical objects have been largely unsuccessful.\cite{mollendal2012rotational, moran2025large, margules2026laboratory, holdren2026rotational, lukova2025millimeter, insausti2025rotational, dhariwal2024origin} To date, the only confirmed chiral molecule detection in the interstellar medium is propylene oxide (2-methyloxirane, \ce{C3H6O}).\cite{mcguire2016discovery} This simple epoxide molecule was observed in the cold extended shell around Sagittarius B2, a molecular cloud abundant in complex organics and dust. The discovery of propylene oxide in this star-forming cloud---in a field of numerous other non-detections of other chiral molecules---prompts the question: What aspects of the reaction network for propylene oxide support sufficient accumulation of this molecule for detection in this object, and can these mechanisms inform further search for other chiral molecules? 

In kinetically controlled molecular clouds, abundances are largely driven by reactivity rates in the gas phase. At present, there are two proposed mechanisms for gas-phase propylene oxide formation, both of which include formation of the propylene oxide cationic intermediate (\ce{C3H6O+}, \POp{}) as the penultimate step.\cite{bodo_formation_2019,hori_theoretical_2022} Once formed, \POp{} is assumed in these models to convert to neutral propylene oxide via electron recombination or charge-transfer reactions. We were therefore motivated to study gas-phase charge-transfer reactions of \POp{} with species that are abundant in Sagittarius B2. Given propylene oxide's relatively low ionization energy (10.22 eV)\cite{NISTWebBook}, only a handful of molecules are capable of neutralizing \POp{}. This astrophysical object is known to be abundant in alcohols such as ethanol and methanol, and other gaseous molecules and dust,\cite{cheung_detection_1968, zuckerman_detection_1975, ball_detection_1970}  but we found the most intriguing candidate reaction partner to be ammonia, \ce{NH3}, which is one of the most abundant molecules in Sagittarius B2.\cite{huttemeister1995multilevel} The low ionization energy of ammonia (evaluated 10.070 eV)\cite{lias2013ionization} means it could play a crucial role in neutralizing \POp{}, creating the conditions needed to support accumulation of stable, neutral propylene oxide in this environment. To evaluate the importance of this reaction, we have measured the charge-transfer rate coefficient for reaction between \POp{} and \ce{NH3}. 

\section{Methods}

The reaction between propylene oxide cation and neutral gaseous ammonia was studied in the glow-discharge ion-trap (GDIT) instrument, which has been described in detail previously.\cite{kisuryna2026development} The instrument consists of a glow-discharge ion source for continuous stable ion generation, an ion guide for ion focusing and transmission, a quadrupole mass filter for mass selection and mass spectrometry analysis, and a linear quadrupole ion trap that serves as a chemical reaction zone. A set of ion optics and ion benders, which also serve as energy filters, are used for ion trajectory manipulation. Individual ion packets released from the trap are detected by two electron-multiplier detectors. The GDIT instrument is currently used for experiments at room temperature.

We previously used GDIT to produce atomic metal ions; however, molecular ions can also be produced in the glow-discharge plasma if suitable precursors are present in the working gas. We produced propylene oxide cations by flowing a diluted mixture of propylene oxide into the glow-discharge chamber. The vapor from a racemic mixture of ($\pm$)-propylene oxide (Thermo Fischer, $\geq$99\%), de-gassed using freeze-pump-thaw, was diluted in a stainless-steel cylinder to a concentration of 5\% propylene oxide in argon (Airgas, 99.999\%). Gas from the mixed cylinder flowed directly into the glow-discharge chamber, controlled at 10 sccm using a mass-flow controller (Alicat). The pressure in the glow-discharge chamber was maintained at 0.7 Torr. The mixed working gas created a stable plasma when sputtering an aluminum cathode, with typical discharge conditions of ca. 1 kV and 5.2 mA. 

Propylene oxide cation is assumed to form in the source region by charge exchange with \ce{Ar+} present in the plasma. The presence of an ion with mass \ce{[C3H6O]+} was confirmed in direct mass-spectrometry experiments through its observation at mass-to-charge ratio \mz{} = 58. We considered the possibility that the ion signal observed at \mz{} = 58 could be due to a structural isomer of propylene oxide: indeed there are many, including propanal, acetone, allyl alcohol, and methyl vinyl ether. Since our starting reagent is commercially obtained propylene oxide, formation of one of these isomers would require a significant structural rearrangement in the ion source overcoming very large energy barriers.\cite{dubnikova_isomerization_2000} In addition, these isomers of \ce{C3H6O} all have lower ionization energies than that for \ce{NH3}, making our observations (below) of charge transfer energetically forbidden. We therefore assume that all chemistry observed here involves the propylene oxide cation, not one of its isomers. Alongside the parent cation, we also observed small fragment signals and minor contaminants from the ambient air such as \ce{CO2} and \ce{O2} in the mass spectrum. Contaminant ions did not enter the ion trap because they were filtered out by the quadrupole mass filter. 

As an improvement since the previous work, we have enhanced the mass resolution by integrating resolution commands in the data-acquisition code that controls the quadrupole mass filter and can now readily resolve peaks that are 1 amu apart. In product-detection mode, we now have a mass resolution $m/\Delta m$ of 112 at \mz{} = 58, an improvement of a factor of 10 over our prior work. This improved mass resolution was essential for the present study, as the ion source produced both propylene oxide cation (\mz{} = 58) and protonated propylene oxide (\mz{} = 59). Distinctly different reaction kinetics were observed when the ion trap was filled with the quadrupole mass filter set to selectively pass each individual species. For experiments carried out in this work, we selectively filled the ion trap with only \mz{} = 58 ions for our study of \POp{} kinetics. The absence of other species in the ion trap---including \mz = 59---was confirmed by mass spectrometry. 

We first measured the ion-trap loss rate in the absence of reaction partners. Following mass selection, \POp{} was held in the ion trap for a range of different trapping times (0.05--2.7 s) and the intensity of the ion signal was monitored as a function of trapping time. The \POp{} signal decays exponentially  through non-reactive processes with a unimolecular rate coefficient of $k_\text{trap}=0.4 \pm 0.2$ s$^{-1}$. Helium is used as the primary bath gas in the ion trap, and $k_\text{trap}$ was found to be independent of ion-trap pressure over a range of $1.74 \times 10^{14}$ to $1.15 \times 10^{15}$ He atoms cm$^{-3}$.

\begin{figure}[h!]
    \centering
    \includegraphics[width=3.25in]{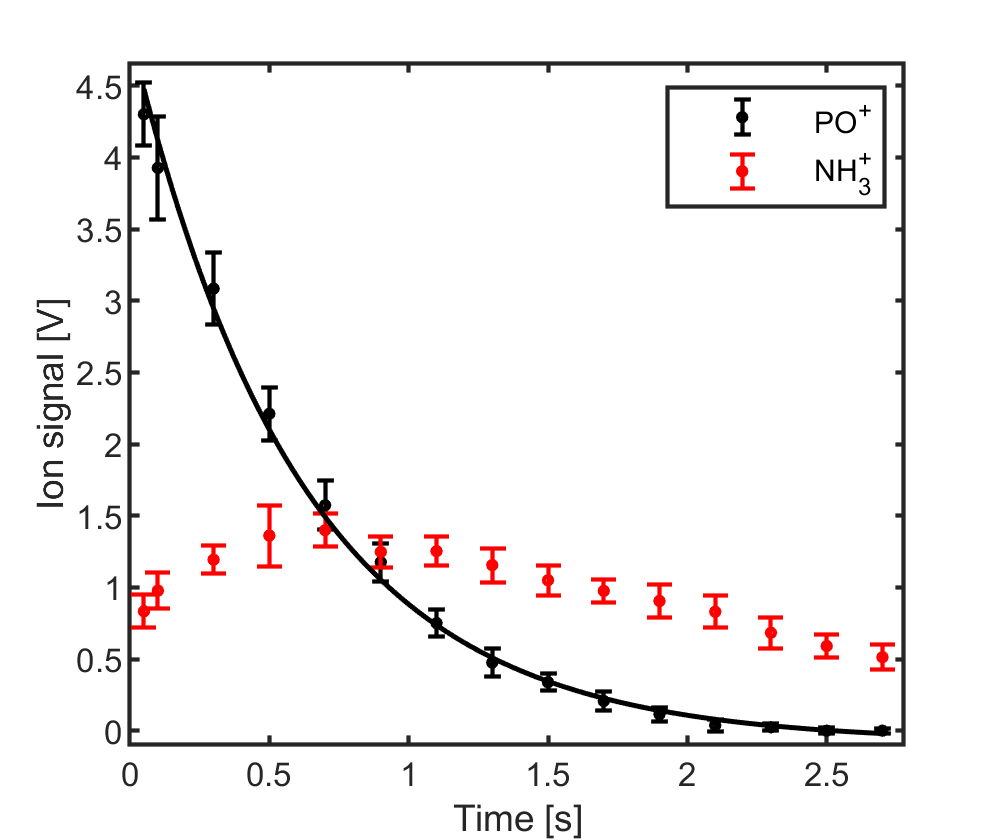}
    \caption{Kinetics profile of the reaction between \ce{C3H6O+} and \ce{NH3}. The black data points show the \mz{} = 58 signal corresponding to \ce{C3H6O+} at various trapping times, while the red data points are the \mz{} = 17 signal from \ce{NH3+}. Each data point is the average of 20 measurements and the error bars reflect the statistical error from random experimental fluctuations. The solid line shows the weighted fit of Eq.~\ref{eqn:int-PO} to the data.}
    \label{kinetics}
\end{figure}

The decay rate of \POp{} signal increased upon the addition of \ce{NH3} at steady-state to the ion trap, indicating a chemical reaction occurred. These experiments were carried out using mixed cylinders of 0.05\% to 0.5\% \ce{NH3} (Matheson, 99.999\%) in He (Airgas, 99.999\%) that were prepared by manometric techniques. The total gas flow into the ion trap was 15 sccm, controlled by two mass-flow controllers (Alicat)---one attached to a pure helium cylinder and the other connected to the mixed \ce{NH3} in helium cylinder. The concentration of \ce{NH3} could be varied in the ion trap by changing the ratio of flow between the two inlet lines, as well as by changing the concentration of the prepared cylinder. The combination of these two strategies resulted in \ce{NH3} concentrations that varied over an order of magnitude. Trapped \POp{} ions reacted with the \ce{NH3} diluted in the bath gas to form products; all charged species remain in the trap and are detected after a pre-set trapping time, corresponding to reaction time. Fig.~\ref{kinetics} shows an example of the exponentially decaying \POp{} signal (black circles) as a function of time with $[\ce{NH3}] = 1.01 \times 10^{12}$ cm$^{-3}$ at a total ion-trap pressure of 20 mTorr. The observed fitted single-exponential decay rate of \POp{} under these conditions is $1.64 \pm 0.08$ s$^{-1}$. 

\begin{figure*}[ht!]
    \centering
    \includegraphics[width=6.25in]{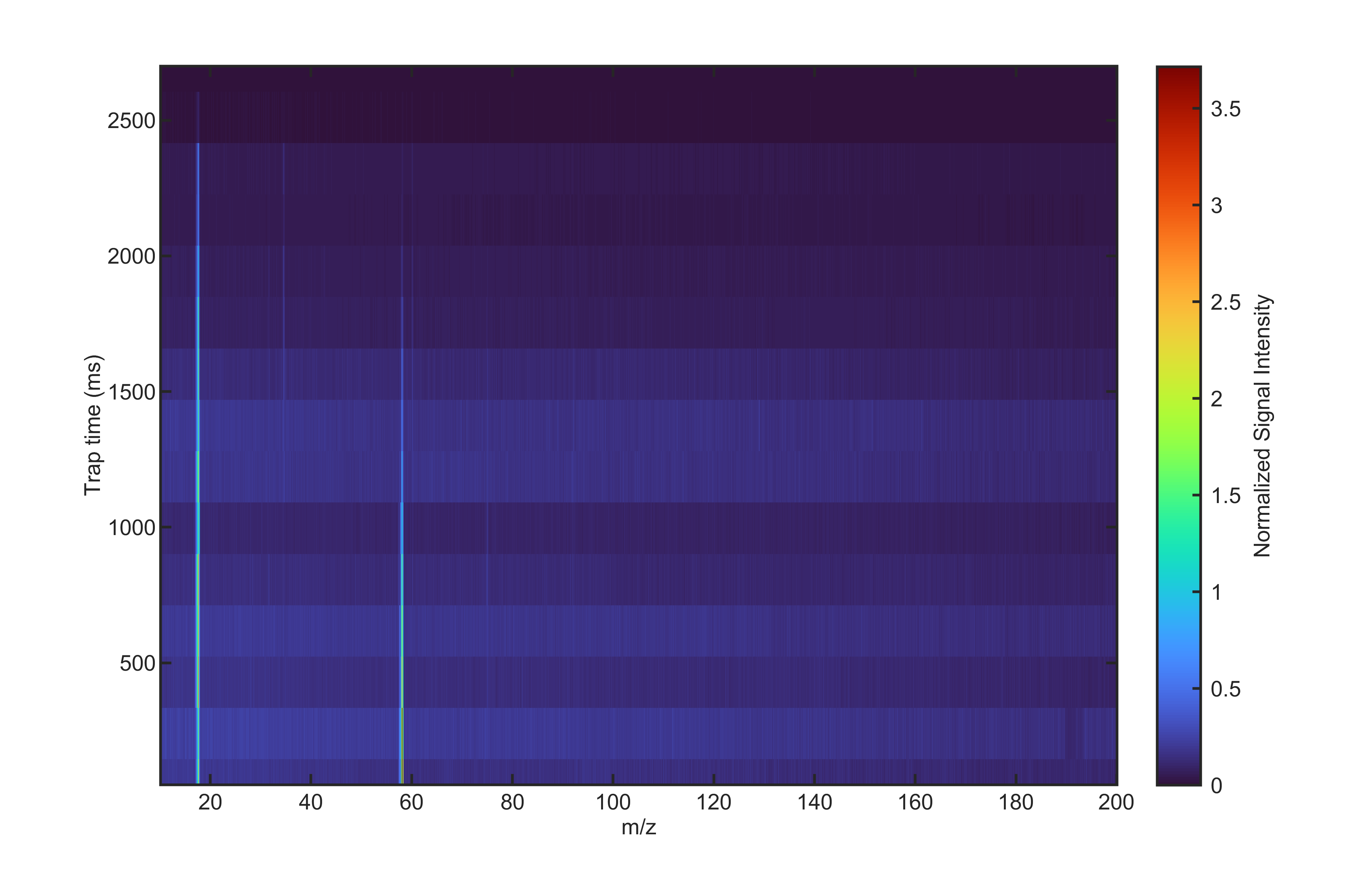}
    \caption{Two-dimensional intensity map of the time-dependent mass spectrum of the reaction of \ce{C3H6O+} with \ce{NH3}. The reactant ion \ce{C3H6O+} appears at \mz{} = 58 and the primary product ion \ce{NH3+} appears at \mz{} = 17. Minor secondary products are observed at \mz{} = 34, 60, 75, and 90. The plot was made was collecting mass spectra at trapping times ranging from 100 ms to 2700 ms in 200 ms intervals. The plot also includes the mass spectra collected at the trapping time of 50 ms.}
    \label{heatmap}
\end{figure*}

We monitored all mass-to-charge ratios within range of our quadrupole mass filter (\mz{} = 10 to 200) as a function of time to observe the time dependence of all charged chemical species in the trap: reactant and products. Fig.~\ref{heatmap} shows the three-dimensional data set (two-dimensional intensity map) of ion signal as a function of trapping time and \mz{} collected at the same conditions as those used in Fig.~\ref{kinetics}. A waterfall representation of the data is also presented in the Supporting Information (SI), Fig.~S1. These figures show the strong ion signal at \mz{} 58 corresponding to \POp{}, along with a significant peak at \mz{} = 17, corresponding to the primary product \ce{NH3+}. The resolution is sufficient to distinguish \ce{NH3+} from other possible products, like \ce{NH4+}. Other minor products are also observed, but at longer trapping times and with much smaller intensities.

Given the relatively sparse mass spectrum, we propose that the majority of the chemistry occurring in the ion trap is driven by a primary charge-transfer step. It is most likely that one of the non-bonding electrons from the \ce{NH3} lone pair is transferred to \POp{}, neutralizing the cation and creating the ammonia radical cation (\ce{NH3+}) with rate coefficient $k_1$: 

\begin{equation}
    \ce{C3H6O+ + NH3 ->[$k_1$] C3H6O + NH3^+}
    \label{rxn:primary}
\end{equation}
\newline
The \ce{NH3+} product forms quickly in hundreds of milliseconds, but eventually peaks and begins to slowly decrease after trapping times of 1 s (Fig.~\ref{kinetics} red symbols). The decrease in \ce{NH3+} signal at long trapping times implies some secondary chemistry is also occurring. The minor masses observed in Fig.~\ref{heatmap} include \mz{} = 75, which has a nominal mass equal to the combination of \ce{C3H6O} and \ce{NH3} (\ce{[(C3H6O)NH3]+}); \mz{} = 34---the ammonia dimer cation (\ce{[(NH3)2]+}); and a higher-order association product at \mz{} = 92. We also observe a minor product at \mz{} = 60, assigned as \ce{[C3H6OH2]+}. All minor species appear only at very long trapping times (more than 1 s) and with very small intensities (less than 9\% the peak intensity for \POp{}). The time-dependent one-dimensional data for these peaks is provided in the SI, Fig.~S2. We assume that the slow loss of \ce{NH3+} at longer trapping times is driven by this secondary chemistry.

We constructed a simple rate law to describe the time-dependent behavior of \POp{} (\ce{C3H6O+}):

\begin{equation}
    \frac{\mathrm{d}[\ce{C3H6O+}]}{\mathrm{d}t} = -k_1 [\ce{C3H6O+}][\ce{NH3}] -k_\text{trap}[\ce{C3H6O+}]
\end{equation}
\newline
where $t$ is the trapping time. In these experiments, the concentration of ammonia is many orders of magnitude larger than the concentration of \POp{} (pseudo-first-order conditions), so we assume that the \POp{} data can be fit to the effective unimolecular rate coefficient $k_1^*=k_1[\ce{NH3}]$, where $k_1^*$ is the pseudo-first-order rate coefficient and the integrated rate law follows

\begin{equation}
    [\ce{C3H6O+}]_t=[\ce{C3H6O+}]_0\exp\left[{-(k_1^*+k_\text{trap})t}\right]
    \label{eqn:int-PO}
\end{equation}
\newline
We fit this model to the \POp{} data (weighted by each point's statistical error) to obtain $k_1^\ast$ from Eq.~\ref{eqn:int-PO}. In all cases, $k_\text{trap}$ was kept fixed at the measured value of 0.4 s$^{-1}$. For the \POp{} data shown in Fig.~\ref{kinetics}, the pseudo-first-order rate coefficient is $k_1^*$ = 1.24 $\pm$ 0.08 s$^{-1}$ after subtracting $k_\text{trap}$. The solid black line in Fig.~\ref{kinetics} is the simulation using Eq.~\ref{eqn:int-PO} showing the good agreement between the model and the data.  The decay of \ce{NH3+} was not observed to trend with \ce{NH3} concentration, implying that it is dominated by complex secondary chemistry that cannot be fit to a simple model.

\begin{figure}[h!]
    \centering
    \includegraphics[width=3.25in]{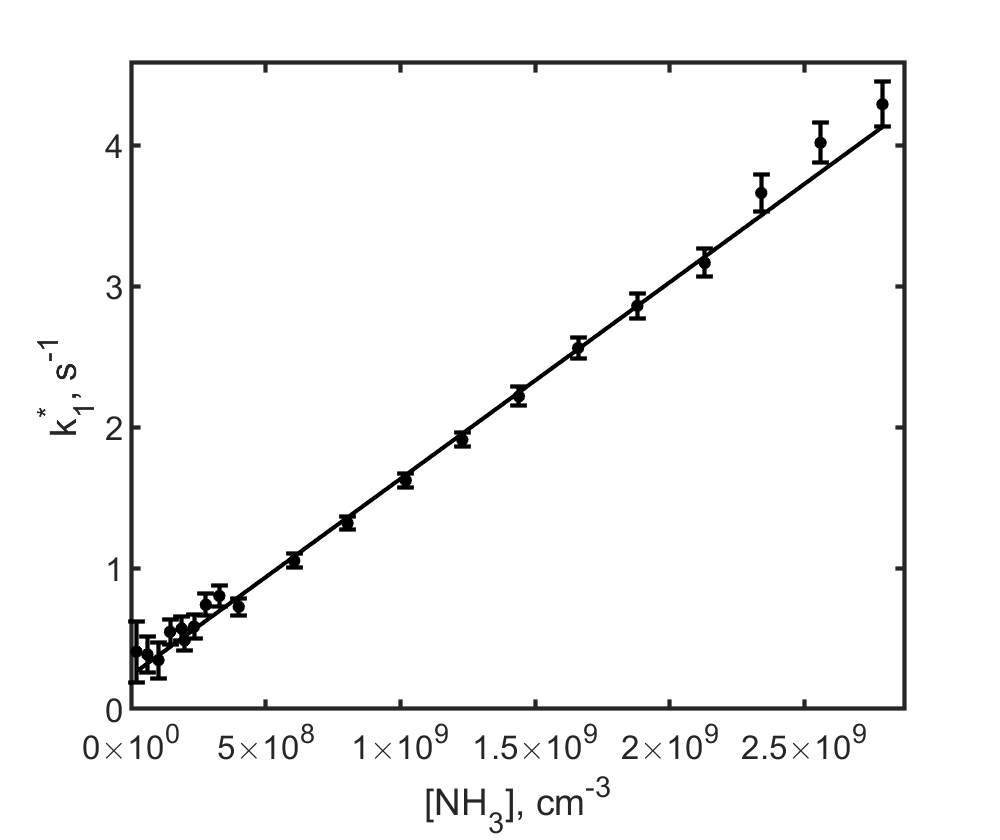}
    \caption{Pseudo-first-order rate coefficient plot showing the dependence of $k_1^\ast$ as a function of ammonia concentration. The data are shown with error bars representing the statistical error in measuring $k_1^\ast$. The solid line shows the weighted linear fit.}
    \label{fig:Conc_Dep}
\end{figure}

To obtain the bimolecular rate coefficient $k_1$, we measured $k_1^\ast$ as a function of [\ce{NH3}], as shown in Fig.~\ref{fig:Conc_Dep}. The data are well described by a straight line, supporting the assumption of pseudo-first-order kinetics. The result of the linear fit yields the charge-transfer rate coefficient $k_1 = (1.39 \pm 0.03)\times 10^{-12} $ cm$^3$ s$^{-1}$ with a $y$-intercept of $0.23\pm0.03$ s$^{-1}$. We tested this system over a range of pressures (19.6 to 36.8 mTorr, for fixed partial pressures of \ce{NH3}) and found that the rate was independent of ion-trap pressure. This supports the conclusion that the reaction proceeds via a pressure-independent charge-transfer reaction under these conditions. 

\section{Discussion}
Our results show that propylene oxide cation (\ce{C3H6O+}) and neutral ammonia (\ce{NH3}) engage in a charge-transfer reaction to form neutral propylene oxide (\ce{C3H6O}) and ammonia radical cation (\ce{NH3+}). We attempted similar measurements with other species that are present in Sagittarius B2, such as carbon monoxide (\ce{CO}), nitrous oxide (\ce{N2O}), methanol (\ce{CH3OH}), hydrogen cyanide (\ce{HCN}), carbon dioxide (\ce{CO2}), oxygen (\ce{O2}), and nitrogen (\ce{N2}) but the presence of any of these species in the reaction trap did not have a measurable effect on the decay of \POp{}. We also did not observe the formation of any products in those experiments, further supporting the assumed lack of reaction. This result is unsurprising because these molecules all have higher ionization energies than that of propylene oxide, and charge transfer would not be energetically favorable. However, this observation also confirms the absence of other two-body mechanisms by which these species could react.

Our results back our proposal that \ce{NH3} likely plays an important role in neutralizing \POp{} to form the neutral propylene oxide observed in the Sagittarius B2 molecular cloud. Indeed, very few species are actually capable of reacting with \POp{} due to its stability and relatively low ionization energy. In addition to electron recombination, Bodo et al. proposed that radicals that have low ionization energies, such as formyl (\ce{HCO}), cyanoethynyl (\ce{C3N}), and hydrocarboxyl (\ce{HOCO}), could be responsible for neutralizing the \POp{} precursor to form propylene oxide. We did not experimentally test the reactivity of these radicals but note that their abundances in the interstellar medium are much lower than that of \ce{NH3}, as would be expected for reactive radical species. For example, the ammonia column density in Sagittarius B2 is reported to be $2.0 \times 10^{16}$ cm$^{-2}$, whereas cyanoethynyl has been detected in TMC-1 with column density of $7 \times 10^{12}$ cm$^{-2}$, and even lower than this in Sagittarius B2.\cite{cheung_detection_1968, friberg_interstellar_1980} Ideally we would use capture theories to predict the rate coefficients for reactions that are not experimentally tractable; however, these theories can fail to predict charge-transfer rate coefficients when short-range interactions are involved.\cite{petralia_strong_2020,tsikritea_capture_2022} We calculated the Average Dipole Orientation (ADO)\cite{su1973theory, su1975parameterization} rate coefficient for \POp{} + \ce{NH3} and obtained $k_{1}(\mathrm{ADO}) = 8 \times 10^{-9}$ cm$^3$ s$^{-1}$, whereas the experimentally measured value is $k_1(\mathrm{expt}) = 1.39 \times 10^{-12} $ cm$^3$ s$^{-1}$, a difference of a factor of nearly $10^4$, clearly indicating for this system that steric factors likely create preferential collision orientations, making simple capture theories insufficient. We can therefore only speculate at the possible importance of \POp{} reactions involving proposed radicals in this environment, but note that the charge-transfer rate coefficients would have to be orders of magnitude greater than that for \ce{NH3} given the significant difference in abundances for these reactions to be important. 

\section{Conclusion}

In conclusion, we propose a new reaction mechanism by which chiral propylene oxide may have been formed in Saggitarius B2: charge transfer between cationic propylene oxide and \ce{NH3}. The rate coefficient for \POp{} + \ce{NH3} charge transfer is slow, observed to be $k_1 = (1.39\pm0.03)\times10^{-12}$ cm$^3$ s$^{-1}$ at room temperature, which is significantly slower than that predicted by capture theory, emphasizing the need for experimentally measured values. Although this reaction is slow, \ce{NH3} is abundant in this molecular cloud, making it a likely reaction partner here. We hypothesize that the neutral propylene oxide may be sufficiently abundant for observation in Sagittarius B2 because of this neutralization pathway. Other chiral molecules of astrophysical interest, such as 2-aminopropionitrile, alanine, and $\alpha$-aminoethanol all have ionization energies even lower than that of propylene oxide (and lower than \ce{NH3}), meaning charge exchange is closed for their corresponding cations.\cite{hrodmarsson2024vacuum, NISTWebBook} It is likely that these chiral candidates exist as cations, not neutrals, in molecular clouds if their formation pathways similarly proceed through their corresponding ion. Polycyclic-aromatic hydrocarbons (PAHs) are likely the only abundant astrochemicals that have ionization energies sufficiently low to undergo electron transfer to the chiral cations.\cite{NISTWebBook} It may therefore be advantageous to search for chiral cation precursors, rather than the neutrals, in regions of space characterized by low electron densities and few PAHs.

\begin{acknowledgement}

This material is based upon instrument and methodology development work supported by the National Science Foundation under Grant No. CHE-2154055. The research for chiral molecules was supported by a Startup grant from the United States--Israel Binational Science Foundation (BSF).

\end{acknowledgement}

\textbf{Supporting Information Available:} Waterfall-plot representation of the data shown in Fig. 2. Time-dependent ion traces for minor secondary products.

\providecommand{\latin}[1]{#1}
\makeatletter
\providecommand{\doi}
  {\begingroup\let\do\@makeother\dospecials
  \catcode`\{=1 \catcode`\}=2 \doi@aux}
\providecommand{\doi@aux}[1]{\endgroup\texttt{#1}}
\makeatother
\providecommand*\mcitethebibliography{\thebibliography}
\csname @ifundefined\endcsname{endmcitethebibliography}  {\let\endmcitethebibliography\endthebibliography}{}


\end{document}